\documentclass[12pt,preprint]{aastex}
\shorttitle{Extended X-ray emission from 4C\,41.17}
\shortauthors{Scharf}

\begin{document}
\title{Extended X-ray emission around 4C\,41.17 at $z=3.8$}

\author{Caleb Scharf}
\affil{Columbia Astrophysics Laboratory, Columbia University, MC5247, 550 
West 120th St., New York, NY 10027, USA.}
\email{caleb@astro.columbia.edu}

\author{Ian Smail}
\affil{Institute for Computational
Cosmology, University of Durham, South Rd, Durham UK}
\email{ian.smail@durham.ac.uk}

\author{Rob Ivison}
\affil{Astronomy Technology Centre,
Royal Observatory Edinburgh, Blackford Hill, Edinburgh UK}
\email{rji@roe.ac.uk}

\author{Richard Bower}
\affil{Institute for Computational
Cosmology, University of Durham, South Rd, Durham UK}
\email{r.g.bower@durham.ac.uk}

\author{Wil van Breugel}
\affil{University of California, Lawrence Livermore National Laboratory, 
Livermore, CA, USA}
\email{wil@igpp.ucllnl.org}

\author{Michiel Reuland}
\affil{Leiden Observatory, Leiden 2300 RA, The Netherlands}
\email{mreuland@igpp.ucllnl.org}

\begin{abstract} We present sensitive, high-resolution, X-ray imaging
from {\sl Chandra} of the high-redshift radio galaxy 4C\,41.17
($z=3.8$).  Our 150-ks {\sl Chandra} exposure detects strong X-ray
emission from a point source coincident with the nucleus of the radio
galaxy.  In addition we identify extended X-ray emission with a
luminosity $\sim 10^{45}$ erg s$^{-1}$ covering a 100\,kpc (15$''$)
diameter region around the radio galaxy.  The extended X-ray emission
follows the general distribution of radio emission in the radio lobes of
this source, and the distribution of a giant Lyman-$\alpha$ emission
line halo, while the spectrum of the X-ray emission is non-thermal and
has a power law index consistent with that of the radio synchrotron.  
We conclude that the X-ray emission is most likely Inverse-Compton
scattering of far-infrared photons from a relativistic electron
population probably associated with past and current activity from the
central object. Assuming an equipartition magnetic field the CMB energy
density at $z=3.8$ can only account for at most 40\% of the
Inverse-Compton emission.  Published submillimeter maps of 4C\,41.17
have detected an apparently extended and extremely luminous far-infrared
emission around the radio galaxy. We demonstrate that this photon
component and its spatial distribution, in combination with the CMB can
reproduce the observed X-ray luminosity. We propose that
photo-ionization by these Inverse-Compton X-ray photons plays a
significant role in this system, and provides a new physical feedback
mechanism to preferentially affect the gas within the most massive halos
at high redshift. This is the highest redshift example of extended X-ray
emission around a radio galaxy currently known and points towards an
extraordinary halo around such systems, where cool dust, relativistic
electrons, neutral and ionized gas, and intense infrared and X-ray
radiation all appear to coexist. \end{abstract}

\keywords{galaxies: active -- galaxies: X-ray -- galaxies: jets -- galaxies: radio galaxies individual: 4C\,41.17 -- radiation mechanisms: non thermal}

\section{Introduction} 

One of the major results from {\sl Chandra} observations of low and
intermediate redshift galaxy clusters is the identification of bubble-like
structures in the X-ray images \citep{mcnamara00}. This phenomenon was
first seen with {\sl ROSAT} in the Perseus and Cygnus-A clusters
~\citep{boehringer93,carilli94b}. These cavities are apparently associated
with energy injection due to nuclear activity from active galactic nuclei
(AGN) in the brightest cluster galaxies.  It appears that powerful radio
jets are particularly effective mechanisms for inputting mechanical energy
to the intracluster medium (ICM).  This activity may influence the
structure and energetics of gas in the highest density regions in the
clusters. The redistribution of gas produced by this injection of energy
may have particularly profound consequences for the rate of cooling of gas
in these regions. This would have an impact on a variety of fields,
including theoretical modeling of galaxy formation -- as large-scale
cooling of gas is a critical route for building the baryonic mass of
galaxies \citep{cole00}.  Studying the influence of AGN on their
surroundings is therefore important for determining the capacity of this
form of feed-back to delay the on-set of gas accretion and star-formation
in massive galaxies at high redshifts. The scaling properties of the ICM
are also best explained by an absence of large amounts of very low entropy
gas in cluster cores \citep{kaiser91,evrard91,mushotzky97}. Finding a
workable mechanism for accomplishing this is at the forefront of cluster
formation studies, e.g.\ \citet{tozzi01a,voit02}, and likely candidates
are heating and cooling of the ICM.

Investigating the possibility that radio sources influence the
thermodynamics of dense gas in their environments is further motivated
by observations which demonstrate that individual, luminous, radio
galaxies at high redshifts ($z>2$) reside in apparently overdense
regions: as traced through excesses of Lyman-break galaxies,
Lyman-$\alpha$ emitters, Extremely Red Objects, X-ray detected AGN or
luminous submillimeter galaxies
\citep{pentericci00,kurk00,chapman00,pentericci02,ivison00}.
Thus the feedback from AGN (and star formation) in bright radio sources
may have had ample opportunity to influence the formation and evolution
of galaxies in regions which are believed to evolve into the cores of
clusters at the present-day.  Studying luminous radio galaxies in
high-density environments at high redshifts therefore provides an
important route to track the changing effects of feedback during the
formation of a massive cluster of galaxies.

Early studies of the X-ray environments of powerful radio galaxies were
limited by the modest spatial resolution of the X-ray observations
\citep{brunetti02}.  However, with arcsecond resolution, high sensitivity
and good spectral resolution from {\sl Chandra} there have been two
published searches for hot gas in the vicinity of radio galaxies and
radio-loud quasars at $z>1$
\citep{fabian01,fabian03a,carilli98,carilli02}. One study detects
extended emission at $z=1.79$ with a mildly favored non-thermal origin
and a remarkable morphology (3C\,294, \citet{fabian03a}) and the other,
with a detection of extended emission at $z=2.16$, has been argued to be
thermal in origin, although the data is somewhat lower signal-to-noise
(PKS 1138-262, \citet{carilli98,carilli02}).

In this paper we present a deep {\sl Chandra} observation of one of the
most distant and powerful high-redshift radio galaxies; 4C\,41.17 at
$z=3.8$.  Recent submillimeter observations of this galaxy using the
SCUBA camera have detected intense, and apparently spatially-resolved,
submm emission from the radio galaxy \citep{ivison00,stevens03}. This
emission is thermal radiation coming from dust grains within the galaxy
which are heated by UV photons from young stars or an AGN (e.g.
~\citet{smail02}).  The far-infrared luminosity of the galaxy is in
excess of 10$^{13}$\,L$_\odot$ and the inferred mass of dust is around
$6\times 10^8$\,M$_\odot$ \citep{stevens03}, both of which suggest
intense star formation is occuring within 4C\,41.17. Further evidence
for extreme activity occuring in the vicinity of the radio galaxy
includes the presence of an extended Lyman-$\alpha$ halo reaching to
$>$100\,kpc (\cite{chambers90,dey97,reuland03}; Figure 5) as well as
spatially-extended [O{\sc ii}] and [O{\sc iii}] emission on similar
scales \citep{vanbreugel02}. ~\citet{dey97} have also shown from
spectropolametric observations that the dominant contribution to the
restframe UV emission in 4C\,41.17 must be due to young stars, with a
formation rate of 140--$1100$h$^{-2}_{50}$ M$_{\odot}$, possibly
triggered by the expansion of the radio source (see also
~\citet{vanbreugel99}).

Strong evidence exists for the immediate environment of 4C\,41.17 being
overdense and consistent with a ``proto-cluster'' region.  In particular,
in parallel to detecting the resolved submm emission from the radio
galaxy, \citet{ivison00} also tentatively identified an overdensity of
bright SCUBA galaxies in a 1-Mpc region around the radio source.  They
interpreted this as evidence for vigorous dust-obscured star formation
and AGN activity during the formation of other massive galaxies clustered
around the radio galaxy.  In addition to this highly obscured, massive
galaxy population, \citet{lacy96} have suggested that there is a modest
excess of Lyman-break galaxies in this region, having found six candidate
$z\sim 3.8$ galaxies in a 1.5-arcmin$^2$ field, roughly centered on
4C\,41.17. They show that this density is slightly higher than that
expected in blank fields, although this comparison is uncertain, and
issues of cosmic variance also make it difficult to assess this
over-density \citep{steidel99}.  We also note that spectroscopy of
candidate Lyman-$\alpha$ emission line objects in this field has so far
confirmed three sources at the redshift of the radio galaxy in a
3.5-arcmin$^2$ field, showing an apparent over-density of Lyman-$\alpha$
emitters in this field ~\citep{vanbreugel02}.
  
Throughout this work we assume a cosmology with $H_0=70$ km s$^{-1}$
Mpc$^{-1}$, $\Omega_0=0.3$, and $\lambda_0=0.7$, consistent with the WMAP
parameters  \citep{spergel03}.  In this cosmology at $z=3.8$, 1$''$
corresponds to 7.1\,kpc, the Universe is 1.6\,Gyrs old and the lookback
time is 88\% of the age of the Universe.

\section{{\sl Chandra} Data Analysis}

The {\sl Chandra} observation of 4C\,41.17 was made during Sept 25--26
2002. 150\,ks of exposure was obtained using the back-illuminated S3
chip in VFAINT mode with two consecutive exposures of $\sim 75$ ksec.
After good-time filtering, 135 ksec of usable exposure remained. The
data were re-processed using the standard {\sc CIAO} tools to exploit
the additional background discrimination of the VFAINT mode telemetry. A
processed image of 4C\,41.17 is presented in Figure 1.  Appropriate
RMF's and ARF's were generated for the primary 4C\,41.17 source location
on the S3 chip. Since the S3 chip is currently CTI-uncalibrated no CTI
correction was made to the data.

This observation detects two components of emission associated with
4C\,41.17.  The first of these is a bright point-source at 06 50 52.07
+41 30 31.17 (J2000)\footnote{This position is accurate to $\pm 0.5''$
and has been corrected to the radio coordinate frame using the
position of a hard X-ray source 40$''$ south of 4C\,41.17 which is
well detected at 1.4\,GHz, the offset between the nominal X-ray and
radio coordinates for this source was: $-0.042$\,s, $-0.09''$.}. 
This  X-ray source lies within 0.5$''$ of the position
of the radio core (which itself lies $\sim 0.5''$ east of B1 in
Figure 2, \citep{bicknell00}) and hence is positionally coincident with
this component to the limit of our astrometric precision.

The second X-ray component is a region of diffuse emission surrounding
this point source (Figures 1, 2 \& 3).  This emission drops away as
$\sim 1/r^2$ to the North-East, but is significantly more extended along
the direction of the radio jet to the South-West. To measure fluxes for
the point-source and extended components we define an aperture for the
point source as a 1$''$ square region which fully encompasses the 90\%
flux-enclosed PSF at the on-axis S3 position.  In contrast, fluxes for
the diffuse component were measured using an elliptical aperture
orientated along the major axis of the 4C\,41.17 emission (see Figure
3).  Background regions were accumulated from (a) 10 blank sky regions
in this S3 data, selected to be close to the aimpoint and therefore less
subject to chip-wide variations, and totaling approximately 4
arcmin$^{2}$ (b) blank sky regions from the available blank-sky files.
The effect of utilizing (a) versus (b) in the following analyses was
negligible given our other uncertainties.

Total count rates were measured for both components in the 0.3--10 keV
bandpass (restframe 1.44--48\,keV). For the diffuse component of 4C\,41.17
a total count of 200 photons was obtained.  The estimated background from
the unresolved X-ray background (XRB) was $75\pm 10$ counts, with an
effective exposure time of 135.113\,ks, yielding a background subtracted
count rate of $9.3\pm 0.9 \times 10^{-4}$\,ct\,s$^{-1}$ (see below for
estimated luminosities).  For the point source, 51 photons were measured
in the $1''\times1''$ aperture. By interpolating the surrounding diffuse
component we have estimated a maximum contribution to the point emission
of $14\pm 4$ photons, the contribution of the unresolved XRB is negligible
for this small region.  The background subtracted point source emission
therefore has a count rate $2.7\pm 0.8\times 10^{-4}$\,ct\,s$^{-1}$.
 
Owing to its smaller photon count we restrict our spectral analysis of the
point source to simply evaluating its hardness ratio (defined as
$(H-S)/(H+S)$ with the energy ranges of the soft band ($S$) of 0.5--2\,keV
and for the hard band ($H$) 2--10\,keV, both in the observed frame).  We
estimate a hardness ratio of $0.00\pm 0.55$ and based on previous surveys we
conclude that this is consistent with the expected characteristics of a
high-$z$ Type II AGN \citep{rosati02,tozzi01b}, although Type I (with
hardness ratios $\sim -0.5$) cannot be ruled out. In estimating the intrinsic
luminosity we have therefore adopted the mean photon index $\Gamma=1.5$,
($N(E)\propto E^{-\Gamma}$), and large intrinsic absorption, n(H{\sc 
i})$\simeq
10^{23}$ cm$^{-2}$ determined by \citet{tozzi01b} for $z>1$ Type-II sources
with similar hardness ratios. Changing these properties by 10--20\% (and
ignoring absorption) has negligible impact on our results.  We determine a
flux for the point source (unabsorbed by Galactic n(H{\sc i})) of $3.4\pm 1.0
\times 10^{-15}$\,erg\,s$^{-1}$\,cm$^{-2}$ (0.3-10 keV) and a luminosity of
$8.8\pm 2.5\times 10^{43}$\,erg\,s$^{-1}$ (0.3-10 keV rest frame), or
$4.6\pm1.3\times 10^{44}$\,erg\,s$^{-1}$ (1.44-48 keV rest frame,
corresponding to the 0.3-10 keV observed band and therefore the least
extrapolation).

There are sufficient photons detected in the extended component for us to
attempt a slightly more sophisticated analysis.  We begin by constructing
a 20 ct per bin, 10 bin spectrum on which conventional spectral analysis
can be performed. We find that this spectrum is inconsistent with thermal
bremmstrahlung but an excellent fit is obtained with a single index power
law (using {\sc XSPEC} \& background subtraction).  The best fit photon
index ($0.3-7$ keV) was $\Gamma = 1.29\pm 0.18$ (90\% confidence limits)
with n(H{\sc i}) fixed at the Galactic value of $1\times 10^{21}$
cm$^{-2}$ (see Figure 4), and reduced $\chi^2=1.3$. For comparison, the
measured hardness ratio is $-0.21\pm 0.31$, also consistent with a hard
spectrum. No satisfactory fit could be obtained with a thermal {\sc
MEKAL} model, $kT$ was constrained as $>97$ keV in the restframe at
$z=3.8$. This is not surprising since even a 10 keV Raymond-Smith
spectrum exhibits relatively little photon count above 4 keV,
corresponding to an observed energy (given $z=3.8$) of 0.83 keV, while we
clearly detect the bulk of flux at higher energies (Figure 4).

Fitting an intrinsically absorbed power law (at $z=3.8$) gave a best fit of
$\Gamma=1.6\pm 0.3$ and n(H{\sc i})$ = 3\pm3 \times 10^{22}$ cm$^{-2}$. We do
not consider this model further as it is difficult to credit such strong
absorption being distributed over such a large volume.  A comparison of the
extended X-ray emission to the North-east and South-west of the point source
shows no significant difference in the spectral properties of these two
regions (see Figure 2).

We also test the spectrum using the preliminary {\sc ACISABS} model 
designed 
to model the modified ACIS soft response due to molecular contamination 
and fitting for energies $0.4-7$ keV ~\citep{chartas02}. With n(H{\sc i}) 
fixed at the Galactic value we obtain a best fit power law of 
$\Gamma=1.57\pm 0.26$ (90\%) and reduced $\chi^2=0.7$. A thermal spectrum
fit with the {\sc ACISABS} response yields a best fit of $42\pm 42$ keV 
(reduced $\chi^2=0.86$).

Using our total count rate for the diffuse emission associated with
4C\,41.17 and assuming the measured photon index of $\Gamma=1.57\pm 0.26$
obtained with the {\sc ACISABS} response and the Galactic n(H{\sc i})
column density we estimate an unabsorbed flux of $1.14^{+0.16}_{-0.11}
(\pm 0.11) \times 10^{-14}$\,erg\,s\,cm$^{-2}$ (0.3-10 keV). The errors
correspond to the uncertainty in the photon index and the Poisson
uncertainty in photon counts respectively and are presented in this form
to allow for the systematic nature of the error propagated from the power
law index.

For our adopted cosmology we therefore estimate an X-ray luminosity of
$L_X=1.5^{+0.4}_{-0.3}(\pm 0.2) \times 10^{45}$\,erg\,s$^{-1}$ (1.44-48
keV rest frame, corresponding to the 0.3-10 keV observed frame) or
$L_X=7.5^{+2.1}_{-1.5} (\pm0.8) \times 10^{44}$\,erg\,s$^{-1}$ (0.3-10
keV rest frame).

We have performed several tests to evaluate the limits on a possible thermal
component in the diffuse emission around 4C\,41.17.  As described above, an
unsatisfactory fit to a thermal spectrum ({\sc MEKAL} model) was obtained 
for
the full spectrum. Subdividing the spectrum between 0.3--2\,keV
(1.44--9.6\,keV rest frame) and 2--10\,keV (9.6-48 keV rest frame), each with
5 spectral bins, also yielded best power-law fits, consistent to within 90\%
errors with each other and with the $\Gamma=1.29$ or $\Gamma=1.57$ ({\sc
ACISABS} applied) overall fit and normalization.  Fitting a joint thermal and
power law spectrum to the full dataset again resulted in an essentially
unconstrained thermal component. Finally, we fit a power law to the $>1$\,keV
data, fixed $\Gamma$ to the best slope ($1.6\pm0.4$) and then fit for the
spectral normalization at $<1$\,keV.  Again, the best-fit normalization was
entirely consistent between these two bands. Very conservatively, the upper
limit to any thermal emission must therefore be equal to, or less than, the
Poisson fluctuation in photon counts in the band 0.3--0.83 (i.e.\ rest frame
1.4--4\,keV, assuming negligible thermal emission from $>4$\,keV). This
yields a count rate limit of $< 4\times
10^{-5}$\,ct\,s$^{-1}$(0.3--0.83\,keV) or an unabsorbed bolometric flux
(assuming a generic 4\,keV plasma) of $\leq 3\times
10^{-16}$\,erg\,s$^{-1}$\,cm$^{-2}$. This corresponds to a bolometric
luminosity limit for a thermal spectrum of $L_X\leq 4.1\times
10^{43}$\,erg\,s$^{-1}$ (i.e.\ $\leq 3$\% of the 1.44-48 keV rest frame
luminosity).  Thus, while we can only rule out the presence of an X-ray
emiting cluster with a luminosity some 40\% of that of the Virgo cluster, it
is clear that the dominant source of the extended X-ray emission is
non-thermal.

In summary: the unresolved X-ray source within 4C\,41.17 has a luminosity
and spectral properties consistent with a heavily obscured, luminous AGN
(Type-II). While the diffuse X-ray component around 4C\,41.17 is most
likely non-thermal in origin, with a power-law spectrum and significant
luminosity. We limit the contribution from any thermal component to less
than $\sim 4\times 10^{43}$\,erg\,s$^{-1}$.

\section{The nature of the diffuse X-ray emission around 4C\,41.17}

In addition to 4C\,41.17, extended X-ray emission has been detected
around a number of moderate and high-redshift radio galaxies and
radio-loud quasars and these exhibit a variety of strength of
correlation between the morphologies in the X-ray and radio bands
\citep{crawford99, chartas00, schwartz00, celotti01, siemiginowska02,
brunetti01, worrall01, carilli02, fabian03a, fabian03b, donahue03,
comastri03}. However, the precise origin of the X-ray emission is
still a matter of debate \citep{harris02}. Where there is a strong
spatial correlation between the emission in the radio and X-ray bands
the prefered mechanism could either be synchrotron emission or
Inverse-Compton (IC) scattering by the relativistic electrons in the
jet plasma of photons from the source itself or the Cosmic Microwave
Background (CMB).  Where there is less apparent spatial correlation
between the two wavebands (possibly due to weaker detections), there is
more ambiguity. In some cases it has been suggested that the X-ray emission 
may be thermal in origin and arise from hot gas in a group or cluster-like 
structure ~\citep{carilli02}. In the case of 3C 294 ~\citep{fabian03a},
the X-ray emission is very clearly much more extended than the radio
structure and exhibits remarkably sharp edge features. ~\citet{fabian03a} 
suggest that this is consistent with IC emission from an older 
population of relativistic electrons generated by previous episodes of
nuclear activity and with confinement by a hot, unseen, ICM.

Further, tentative support for an IC mechanism operating in some
high-redshift radio galaxies studied in the X-ray and radio bands comes
from the apparent evolution of the $L_X/L_R$ ratios.  IC scattering from
the CMB implies that the ratio of X-ray to radio luminosities should scale
as $(1+z)^4$ (see Equation 1 below). Using published X-ray luminosities
(see references above) and radio powers ~\citep{herbig92} we find
$L_X/L_R=0.70$ (4C\,41.17, $z=3.798$), 0.21 (PKS\,1138$-$262, $z=2.156$),
0.11 (3C\,9, $z=2.012$), and 0.08 (3C\,219, $z=0.174$). These ratios are
clearly subject to large uncertainties, such as the modeling of the X-ray
emission, magnetic fields, and the presence of local far-infrared (FIR)
photon sources. However, a general trend for increasing $L_X/L_R$ with
redshift appears to be present, although a poor fit to the expected
$(1+z)^4$ scaling.

In the case of 4C\,41.17, several pieces of evidence also point to a
non-thermal IC origin of the diffuse emission. First, the restframe
surface brightness of the emission is $(1+z)^4 I_x\sim
10^{-11}$\,erg\,s$^{-1}$\,cm$^{-2}$\,arcmin$^{-2}$. This is a factor $\sim
100$ higher than ICM thermal emission in the brightest local galaxy
clusters. IC scattering of CMB photons on the other hand has almost no
redshift dependence owing to the cancellation of surface brightness
dimming by the corresponding increase in energy density of the CMB with
redshift (although variations in magnetic field strength come into play).

Second, the measured spectrum of the X-ray emission is a power law, with
an energy spectral index $\alpha \equiv 1-\Gamma=-0.57\pm0.26$. As
discussed by ~\citet{chambers90} the integrated radio spectrum of
4C\,41.17 has a low frequency index of $\alpha \simeq -0.83$ ($\leq 200$
MHz rest frame), where synchrotron losses have a longer timescale, and an
index of $\alpha \simeq -1.33$ for higher frequencies (200MHz-48GHz rest
frame). If the X-ray emission is IC scattering then a large fraction of
the flux must arise from electrons with synchrotron frequencies $<200$MHz
(see Equation 2 below) and should exhibit the same spectral slope
~\citep{felten66}. Given the uncertainties in both the X-ray and radio
spectral fits it appears that the energy spectral indices are indeed
consistent -- as would be expected if the X-ray and radio photons
originated from the same particle population.

Thirdly, the overall morphology of the X-ray emission is indeed correlated
with the radio maps (Figure 2), especially if one notes that the observed
radio emission is at $\sim 7.5$ GHz rest frame. Since the radio spectrum
is steep in this range ($\alpha \sim -1.3$) only the highest surface
brightness features are seen in the maps. Moreover, since the radio source
spectra steepen away from the hotspot towards the AGN, the radio
morphology at low frequencies is somewhat different from that at higher
frequencies, with relatively bright `bridges' and fainter hotspots (e.g.  
Cygnus-A). A large abundance of low energy electrons in the bridges would
result in increased IC losses and thus brighter X-ray emission from these
structures as compared to the hotspots (modulo the photon field
distribution, see below). The primary electron population for IC
scattering of the CMB photons must have $\gamma\sim 1000$, or if we are
seeing IC scattering of local FIR photons then the electrons will have
$\gamma\sim 100$.  By contrast, the electrons responsible for the bulk of
the observed radio synchrotron emission (5-25 GHz rest frame) have
$\gamma\sim 5000$ \citep{brunetti02}.  The general, but not
point-by-point, correlation of radio and X-ray emission (Figures 2 \& 5)
is therefore not unexpected and offers clues to the electron distribution
and AGN activity. We discuss this in more detail below.

Finally, although our X-ray data are currently too noisy for a definitive
measurement, the surface brightness distribution appears to be relatively
flat but with quite well delineated edges (e.g. Figure 1). As 
discussed by ~\citet{fabian03a} for the case of 3C\,294, this suggests 
confinement of an electron population by an external, hot, medium -- such 
as that already posited for 4C\,41.17 (~\citet{carilli94} and see \S4).

While other, non-IC, non-thermal sources of X-ray emission cannot be entirely
ruled out, all such sources will be dimmed by $(1+z)^4$, a
factor of 530 at $z=3.8$, putting serious constraints on possible
emission mechanisms.  We suggest therefore that the most likely source
for the diffuse X-ray emission around 4C\,41.17 is IC scattering of CMB/FIR 
photons.  We now discuss this mechanism in more detail,
as well as the probable origin of the photons: either the CMB alone or 
with a contribution from the known luminous FIR source.

The relative luminosity of the radio synchrotron and IC X-ray emission
from scattering off a photon field is given by:

\begin{equation}
\frac{L_R}{L_X}=\frac {B^2/8\pi}{\rho}\;\;\; ,
\end{equation}

where $B$ is the magnetic field strength in the plasma and $\rho$ is the
energy density of the scattering photons. For the CMB, at a redshift $z$,
$\rho_{CMB}=7.56\times 10^{-15} T_0^4 (1+z)^4$\,erg\,cm$^{-3}$
 \citep{jones65,felten66,schwartz02a,schwartz02b}. At $z=3.8$ with
$T_0=2.728$K \citep{fixsen96} $\rho_{CMB}=2.2\times
10^{-10}$\,erg\,cm$^{-3}$.

The relationship between IC scattered energies, $E$, and the synchrotron 
frequencies emitted by electrons of a given energy is:

\begin{equation}
\left< E\right> (eV) \approx 0.9\times 10^2 \frac {T}{B (\mu G)} \nu_{s} 
(MHz)\;\;\; ,
\end{equation}

where $T$ is the characteristic photon temperature and $\nu_s$ is the 
synchrotron radio frequency ~\citep{felten66}.

The radio luminosity of 4C\,41.17 integrated over all components is
$L_R=1.28 \times 10^{46}$\,erg\,s$^{-1}$ (from \citet{chambers90} and
converted to our adopted cosmology). In the following calculations we
adopt a constant spectral index from 10-10,000 MHz of $\alpha =-1.33$
($S_{\nu}=k \nu^{\alpha}$). This is conservative in the sense that it will
slightly over-predict the low frequency luminosity and therefore slightly
over-predict the expected IC X-ray emission. In Figure 5 we plot the IC
X-ray luminosity predicted from the known radio spectrum of 4C\,41.17 and
Equations 1 and 2 (in a 1.44-48 keV rest frame energy band corresponding
to the observed 0.3-10 counts and therefore requiring minimal
extrapolation). The predicted $L_x$ is shown for both CMB photons alone
and for 3 cases in which the local mean FIR energy density over the
emitting volume is 1, 2, and 3 times the CMB energy density.

 ~\citet{carilli94} determine an equipartition field of $B=$50--80$\mu$G
(their Appendix A, adjusted to our assumed cosmology) in the low surface
brightness regions away from the radio hotspots. Figure 5 demonstrates
that a significant contribution from the local FIR photons is necessary
to produce the observed $L_x$ if the equipartition $B$ field is correct.

As discussed in \S1, 4C\,41.17 is one of the most luminous sub-mm sources
known. The galaxy is detected at both 450\,$\mu$m and 850\,$\mu$m with fluxes
of 35.3\,mJy and 11.0\,mJy respectively \citet{ivison00}, implying a graybody
temperature of $47\pm10$ K and $L_{FIR}=3.2^{+3.2}_{-1.7} \times
10^{13}$L$_{\odot}$ ($1.22^{+1.22}_{-0.65}\times 10^{47}$\,erg\,s$^{-1}$) for
our assumed cosmology. We note that this exceeds the integrated
radio-synchrotron luminosity by a factor $\sim 15$, showing that the sub-mm
emission arises in a separate component of dust-reprocessed radiation from
vigorous star formation or an AGN \citep{ivison00}.

If the local FIR emission is point-like and isotropic then the energy
density at a distance $R$ from the source is $\rho_{FIR}=3L/4\pi cR^2$.
For the observed $L_{FIR}$ then the local photon energy density would
exceed that of the CMB, $\rho_{FIR}>\rho_{CMB}$, out to a distance of
13\,kpc from the source. Clearly then the local FIR photons can play a
significant role in the observed IC emission and may provide the
necessary boost in $L_x$ required by Figure 5.

There is also evidence that the FIR source in 4C\,41.17 is spatially
extended. \citet{ivison00} claim that the thermal dust emission from
4C\,41.17 is partially resolved on a scale of 50\,kpc (7$''$, FWHM),
after correcting for the SCUBA beam (Figure 2; see also
\citet{stevens03}). This implies a physical extent of the submm emitting
region of $\sim 50$\,kpc.  We adopt a very crude model for the dust cloud
of a sphere of uniform emissivity and diameter $D_{cloud}$, this yields a
mean FIR energy density of $\simeq 9L_{FIR}/2\pi c D_{cloud}^2$ within
its volume.  In this case the mean local FIR energy density in the cloud
exceeds that of the CMB if $D_{cloud}$ is {\em smaller} than
$53^{+22}_{-17}$\,kpc (errors from uncertainties in $L_{FIR}$). A
slightly more realistic model has the relativistic plasma modeled as a
cylindrical volume of length $\sim 100$\,kpc, embedded in the cloud, this
would then require that $D_{cloud}$ must be {\it smaller} than $\sim
60$\,kpc for the volume averaged energy density to be equal to or greater
than that of the CMB over the entire emitting region, $\rho_{FIR}\geq
\rho_{CMB}$.

Since this is comparable to the claimed size of the dust emission in this
system from \citet{ivison00} it indicates that local FIR emission could
plausibly make up the apparent shortfall in photons seen in Figure 5,
although these estimates are subject to significant uncertainties and are
designed to be illustrative rather than definitive.  We note that the
relatively flat X-ray emission profile in 4C\,41.17 also suggests that
the dominant source of local FIR photons arises in an extended component.  
However, the tentative evidence on the North-Eastern side for a $\sim
1/r^2$ drop-off (\S2) suggests that the local FIR emission in that
direction could have a more compact configuration.

We conclude that it appears necessary and likely inevitable that local FIR
photons play a role in the IC X-ray emission (Figure 5). However, the
inherent difficulties in constraining magnetic field strengths and the
unknown true distribution of the electron population make more detailed
estimates very difficult.

\section{4C\,41.17: a forming massive galaxy in a
proto-cluster}

Early indications of the presence of a high-density gas environment
around 4C\,41.17 came from the detection of a strong and spatially
variable rotation measure in polarimetry mapping of the various radio
structures in the source \citep{carilli94}.  These rotation measures
indicate the presence of large column densities of magnetized, hot
gas.  \citet{carilli94} estimate a total gas mass of $\geq
10^{11}$\,M$_\odot$ (in our cosmology) and suggest that this material
arises from an X-ray type ``cooling flow'' in a dense cluster
environment around 4C\,41.17. The results presented here do not rule out
this scenario.  Our limit of $\sim 4\times 10^{43}$ erg s$^{-1}$ on
thermal emission is still compatible with a much more spatially extended
thermal component which may be accreting into the central potential and
which would be surface brightness dimmed by $1/(1+z)^4$.

In an effort to further understand the distribution of matter in 
4C\,41.17 we make use here of a very deep Lyman-$\alpha$ image produced 
with a custom-built, high throughput interference filter
with a 65~\AA\ bandpass centered at the redshifted Lyman-$\alpha$ line
at 5839~\AA\ (for $z = 3.798$) with the Echellette Spectrograph and
Imager ~\citep{sheinis00} at the Cassegrain focus of the Keck II 10m
telescope (these data have been previously discussed in
~\citet{vanbreugel02} and \citet{reuland03}).  The data were obtained
during photometric conditions and good seeing (FWHM = 0.57$''$). With a
total exposure time of 27.6\,ks this Lyman-$\alpha$ image is the most
sensitive obtained to date and reaches a surface brightness of $8.0
\times 10^{-19}$\,erg\,$s^{-1}$\,cm$^{-2}$\,arcsec$^{-2}$ (3$\sigma$
limit in a 2.0$''$ diameter aperture).  A broad-band $R$ image, obtained
as part of a multi-band spectral energy distribution study of the
4C\,41.17 field \citep{reuland03} was scaled and subtracted from the
Lyman-$\alpha$ image to construct the pure emission-line image shown in
Figure 6. In addition,  extended [O{\sc ii}]\,$\lambda$3727 and [O{\sc
iii}]\,$\lambda$5007 emission has been detected, confirming that the
Lyman-$\alpha$ arises from an ionized medium and not from scattering off
a neutral gas. With a linear dimension of $\sim 150$\,kpc at the
detection limit, the 4C\,41.17 nebula is the largest presently known
 \citep{reuland03}.

The Lyman-$\alpha$ halo shows distinct correlation features with the
X-ray and radio maps, including several spur-like Lyman-$\alpha$
features, which also appear on slightly larger scales in the X-ray
emission. The IC X-ray model as described in \S3 might suggest that the
X-ray emission would tend to be anti-correlated with Lyman-$\alpha$
emitting regions owing to the propensity of radio jets to evacuate gas
from their volume. Indeed this would appear to match reasonably well the
situation seen around the South-Western jet, A, in Figure 6.  The
Lyman-$\alpha$ spurs to the North and East are are then anomalous,
unless we consider an additional mechanism for producing a population of
relativistic electrons correlated with the ionized Lyman-$\alpha$ gas.

The halo of a massive proto-galaxy is likely to consist of a multiphase
medium of cool, $\leq 10^4$\,K, clouds or filaments embedded in a
virialized medium at $10^6$--$10^7$\,K (Rees 1989).  In such a
multi-phase medium the passage of a bow shock driven by an expanding
radio source, or by starburst activity, may stimulate a range of
behaviour.  The densest patches are likely to be relatively unaffected
by the passage of the shock, although they may collapse due to the
increased pressure of the gas surrounding them.  In contrast, the
intermediate-density clouds will be shocked and ionized, cooling back
via Lyman-$\alpha$ emission. These shocks may also re-accelerate any
older, cooler relativistic electrons caught in the magnetic fields
around the radio galaxy.  In this picture, there is a direct physical
correspondance between regions of Lyman-$\alpha$ emission, relativistic
electrons and X-ray emission as the electrons cool via IC scattering off
CMB and local FIR photons. The need for a more localized electron
``re-heating'' is also suggested by comparing the $\sim 100$\,Myr
crossing time of the 4C\,41.17 system to the IC cooling time of the
electrons (assuming CMB photons only): $2.1\times
10^{12}/[\gamma(1+z)^4]$\,yr \citep{schwartz02b}, which is $\sim 4$\,Myr
for $\gamma=1000$ at $z=3.8$.  Further support comes from the detection
of high-velocity outflowing gas around 4C\,41.17. Both the [O{\sc ii}]
and Lyman-$\alpha$ emission lines exhibit large blue-shifted velocities
($\sim 600$--900 km s$^{-1}$) along the direction of the principle radio
axis and South-Western Lyman-$\alpha$ filament, together with
Lyman-$\alpha$ velocity widths in this filament as high as $\sim
900$--1600 km s $^{-1}$ ~\citep{vanbreugel02}. Along this axis it seems
likely therefore that shock heating of the gas will play a major role in
powering the Lyman$-\alpha$ emission, and potentially contributing to
the accelerated electron population responsible for the IC X-ray
emission. Away from this axis however the situation is less clear, since
there is still widely distributed IC X-ray emission with correlated and
quite uniform, Lyman-$\alpha$ emission.

\section{Feedback and galaxy growth}

Finally, and most excitingly, it may be possible for the X-ray
emission to produce Lyman-$\alpha$ emission through local photo-ionization
of gas.  The ionization parameter can be estimated to order of magnitude
as: $\xi \sim (10^{45}{\rm erg s^{-1}})/nl^2$, where $n$ is the mean gas
density (cm$^{-3}$) and $l$ is the physical scale of the gas distribution.
Taking $l\sim 100$ kpc then $\xi \sim 1/n100$. We know that [O {\sc ii}]
and [O {\sc iii}] species exist in the halo gas \citep{vanbreugel02},
consequently $\xi$ must be in the range 1--10 \citep{kallman82}, and
probably towards the lower end in order for the recombination
Lyman-$\alpha$ emission to be seen. Very crudely then, if the typical halo
gas density is {\it less} than $\sim 0.001$ cm$^{-3}$ (as suggested from
the rotation measure, \citet{carilli94}) the observed X-ray emission would
photoionize the gas to much higher species than observed. However, in the
denser regions it would appear that direct photoionization by the IC X-ray
emission is another viable, if not inescapable, mechanism for creating the
ionized Lyman-$\alpha$ halo. This mechanism could operate either in
addition to, or independently of the shock driven ionization, depending on
the location relative to the principle jet/out-flow axis as discussed
above. At the present moment it is impossible to differentiate between
these scenarios, but future optical and near-IR IFU observations of the
spatial distribution of a range of emission lines in the halo of 4C\,41.17
will be a powerful technique to investigate the internal physics of this
galaxy.

If it can be demonstrated that it operates, this new IC X-ray
photoionization mechanism presents the AGN within 4C\,41.17 with another
route to exert feedback on the surrounding gas halo on a larger scale
than could be directly surmised from the properties of the radio jets.  
Moreover, this feedback mechanism may continue to operate after the
radio activity has ceased to be directly detectable. Combined with the
mechanical energy input of the relativistic electrons, which is likely
responsible for the X-ray cavities seen in low-z clusters
~\citep{fabian03a}, these represent two major pathways for supermassive
black holes to exert significant feedback on scales extending beyond
their host galaxies.

Perhaps the most important feature of this mechanism is that it is 
expected to be most effective in the most massive halos at the
highest redshifts: those which have been able to build a supermassive
black hole large enough to power powerful radio jets. The mass-specific
nature and redshift dependency of this feedback mechanism may provide
a new avenue for theoretical attempts to model the formation and growth
of the most massive galaxies and clusters seen in the local Universe
\citep{benson03,wu00}.

\section{Conclusion}

We have detected extended X-ray emission around the luminous radio galaxy
4C\,41.17 at $z=3.8$.  The spectral properties of the diffuse X-ray
emission are best described by a power-law model, with a spectral index
which is consistent with that seen in the radio waveband.  Moreover, the
morphology of the X-ray emission correlates well with the general
distribution of radio emission. We propose that Inverse Compton scattering
is the mostly likely origin of the X-ray emission and we limit any
contribution from thermal emission by hot gas to less than $4\times
10^{43}$\,erg\,s$^{-1}$.

We quantify our discussion of IC scattering around 4C\,41.17 and conclude
that if the equipartition estimates of magnetic field strength are valid
then the observed X-ray emission is due to the scattering of both CMB and
local FIR photons. The detection of the galaxy as a very bright submm
source supports this hypothesis.  We compare two simple geometrical
models for the thermal dust emission around 4C\,41.17. The X-ray
requirements for the mean FIR energy density in the plasma volume are
that $\rho_{FIR}=1-3\times \rho_{CMB}$. The constraints that this imposes
on the volume of the FIR emitting region are consistent with the crude,
direct measurements of the size of the galaxy in the submm waveband,
which imply that the dust emission is extended over $\sim 50$\,kpc scales
(although subject to considerable uncertainty).

We demonstrate that IC X-ray emission, enabled by the enhanced CMB at
high redshift and the local FIR luminosity of this highly-obscured,
active system, can readily ionize the gas in the halo of this galaxy.  
While it is unlikely that this is the sole mechanism by which the
Lyman-$\alpha$ halo is produced, it seems inevitable that it plays a
significant role, especially away from the large out-flow velocities
seen along the principle radio axis. It could also be very efficient at
ionizing the less dense, outer halo.  Importantly, if the observed
Lyman-$\alpha$ emission is recombination following photo-ionization then
this gas could also have a significantly lower temperature, and shorter
cooling time.

IC X-ray emission from high-$z$ supernovae has been suggested as a
reionization mechanism in the early universe ~\citep{oh01}, with
profound implications for the formation of stars and galaxies. On the
basis of our results presented here we are compelled to add that IC
X-ray emission from early AGN activity should also be an important
feedback mechanism -- especially given that the diffuse IC emission from
the CMB can easily exceed that directly due to accretion onto the
central black hole, and that its diffuse nature permeates the halo
volume more thoroughly. Moreover, this form of feedback will operate
preferentially in the most massive halos, those which are able to host
powerful radio galaxies.  This mechanism, together with the mechanical
energy injected by the AGN via relativistic electrons may therefore
address the fundamental problem of overcooling in models of the 
formation of massive
galaxies and clusters in the local universe \citep{benson03,wu00}. As we
have shown, 4C\,41.17 exhibits three of the main features currently
expected to regulate star formation in massive galaxies
\citep{benson03}: an ordered magnetic field (detected as a strong
rotation measure) to allow conduction of heat from the outskirts of the
halo into the centre and so suppress the cooling of gas; AGN-driven
feedback in the form of the mechanical input from the radio jets and
lobes; and outflows associated with visible and obscured massive star
formation.  However, it is intriguing that even this combination of
three mechanisms apparently hasn't {\it yet} been able to suppress the
star formation in this massive galaxy (Dey et al.\ 1997; Ivison et al.\
2001).

Finally, we suggest that the combination of intense, but obscured,
activity in 4C\,41.17, coupled with its high redshift, and hence the
strong increase in the energy density of the CMB, has aided in our
identification of the X-ray halo around this galaxy.  If correct this
would imply that large halos, containing large populations of low energy
electrons (tracers of past radio activity), ordered magnetic fields and
significant quantities of ionized gas, may be common around all massive
galaxies at high redshifts, $z\geq 1$. If the IC X-ray emission plays an
important photo-ionization role this would also provide an additional
mechanism for switching on Lyman-$\alpha$ halos at high-$z$ by radio
activity and the CMB/FIR photons alone. Clearly the rapid decline in the
energy density in the CMB, coupled with a similar strong decline in the
typical FIR luminosities of active galaxies at $z<2$
\citep{archibald01,page01} mean that IC scattered X-ray halos should
rapidly decrease in intensity at lower redshifts.

Together with existing data from other wavebands these {\sl Chandra}
results point towards an extraordinary environment in systems like
4C\,41.17. The central AGN is periodically injecting high energy
electrons and energy through the length of the medium in which it is
embedded. This medium is truly multi-phase, consisting of $\sim 50$K
dust distributed over 10--20\% of the 100--150 kpc wide system volume,
co-existing with the largest Lyman-$\alpha$ halo known along with
cooler dust, magnetic fields, intense UV flux from massive star
formation, and flooded by X-ray photons.

As the construction site of a massive galaxy, and almost certainly what
will become the core of a massive cluster of galaxies, this system
directly illustrates the extent to which astrophysics helps mold the
nature of the relatively quiet cluster environments we see today.  The
radio observations of 4C\,41.17 suggest that it is surrounded by a large
mass of gas, which would be a significant component in the ICM of any
present-day cluster, although it has yet to be heated or compressed
sufficiently at $z=3.8$ to be visible in the X-ray waveband.  The
coincidence of this phase of growth with the wide range of energetic
phenomenae we observe in 4C\,41.17 suggest that these phenomenae may
play a role in setting the final physical characteristics of the ICM --
such as establishing a minimum entropy for gas in clusters at the
present-day.  Future X-ray observations of other high-z radio galaxies
will lead to a more unified picture of energetics and feedback
mechanisms in these systems.

\acknowledgements 

We acknowledge Jason Stevens and Jim Dunlop for their work on the submm
properties of 4C\,41.17.  We also acknowledge useful discussions with
David Helfand and Frits Paerels. This work was made possible by the
exceptional capabilities of the NASA/{\sl Chandra} observatory, and the
dedicated support of the CXC. C.A.S. acknowledges the support of NASA
grant NAG5-6035, and NASA/{\sl Chandra} grant SA0 G02-3167X, I.R.S.
acknowledges support from the Royal Society and Leverhulme Trust.  
R.G.B. acknowledges support from PPARC and the Leverhulme Trust. The
work by W.v.B. and M.R. was performed under the auspices of the U.S.
Department of Energy by the University of California, Lawrence Livermore
National Laboratory under contract No. W-7405-Eng-48.  W.v.B. also
acknowledges NASA grants GO~5940, 6608 and 8183 in support of high
redshift radio galaxy research with HST. Finally, we thank the referee
D. Harris for comments which have substantially improved this manuscript.

\begin{figure} \caption{$25''\times 25''$ image of 4C\,41.17 in the
0.3-5 keV band, smoothed using a Gaussian of width $0.5''$ to illustrate
the strong, unresolved nuclear point source and extended low-surface
brightness emission on a scale of $15''$ or 100 kpc (image credit H.
Jessop). North is up, East to the left. }

\plotone{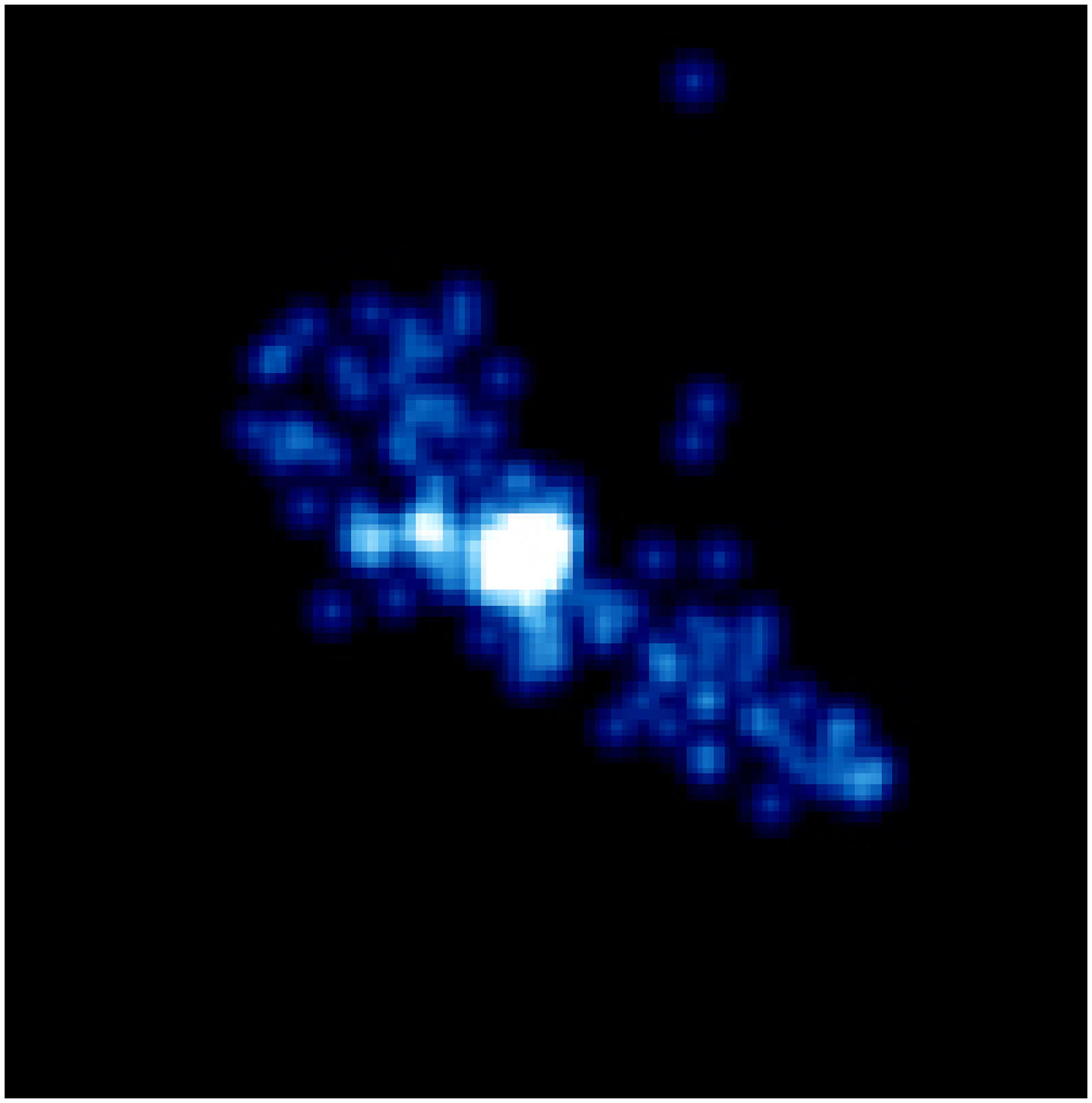}
\end{figure}

\begin{figure} \caption{ Three comparisons of 4C\,41.17 in different
wavebands. All three panels show a $20''\times 20''$ field with North top
and East to the left.  The left-hand panel shows a true color {\sl
Chandra} X-ray image of the source coded as 0.2--1.5\,keV (red),
1.5--2.5\,keV (green) and 2.5--8.0\,keV (blue), with the 850$\mu$m SCUBA
map overlayed (Ivison et al.\ 2000). The lowermost submm contour is at
2.6 mJy per beam, subsequent contours are at 2.9, 3.4, 4.1, and 5.3 mJy
per beam, to best illustrate the morphology.  The submillimeter source is
marginally resolved by the 15$''$ FWHM SCUBA beam (Stevens et al.\ 2003).  
The central panel is the {\sl Chandra} 0.5--2.0\,keV soft-band image
smoothed to match the beam-size of the overlayed 1.4\,GHz VLA map. The
X-ray gray-scale runs from 0.08-0.8 counts per $0.5''$ pixel, the 1.4GHz
contours are at 0.3, 1.9, 10.9, 21.2, 37.5, 63.3 and 104.2 mJy per pixel.  
The rightmost panel shows the 2--10\,keV hard-band image in grayscale
(0.08-0.8 ct per $0.5''$ pixel) overlayed with the 4.9GHz contours (0.08,
0.5, 1.1, 2.2, 3.8 mJy per pixel). We label on this figure the various
radio components using the naming scheme from \citet{chambers90}.
Although correlated, the outermost radio peaks do not coincide with the
major components of the rather flat X-ray emission -- consistent with an
origin in different electron populations (\S3) and with a contribution
from a spatially uniform photon source population such as the CMB. }
\plotone{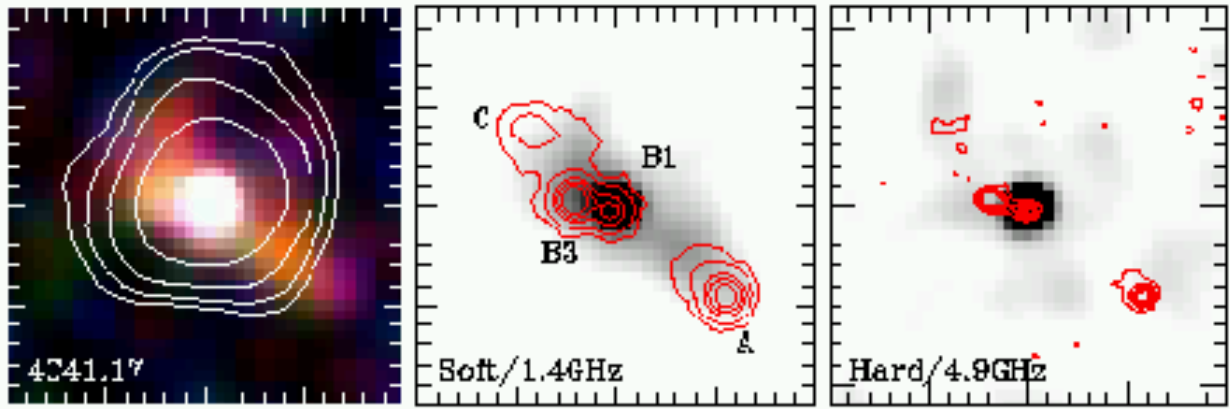} \end{figure}

\begin{figure} \caption{Left panel is 20$''\times$20$''$ region showing
adaptively smoothed 0.5--2 keV {\sl Chandra} data (lowermost contour
0.05 ct per $0.5''$ pixel, linear increments of 0.1 ct per pixel)
overlaid on 1.4GHz radio map grayscale (logarithmic intensity scaling,
$0.01-2$ mJy). Right panel is 40$''\times$40$''$ region showing 0.5--2
keV X-ray data binned to 1$''$ pixels (grayscale 0-6 cts per 
pixel) commensurate with the on-axis {\sl
Chandra} PSF. The region used to extract counts for the diffuse emission
is shown as an ellipse in the right panel.} \plotone{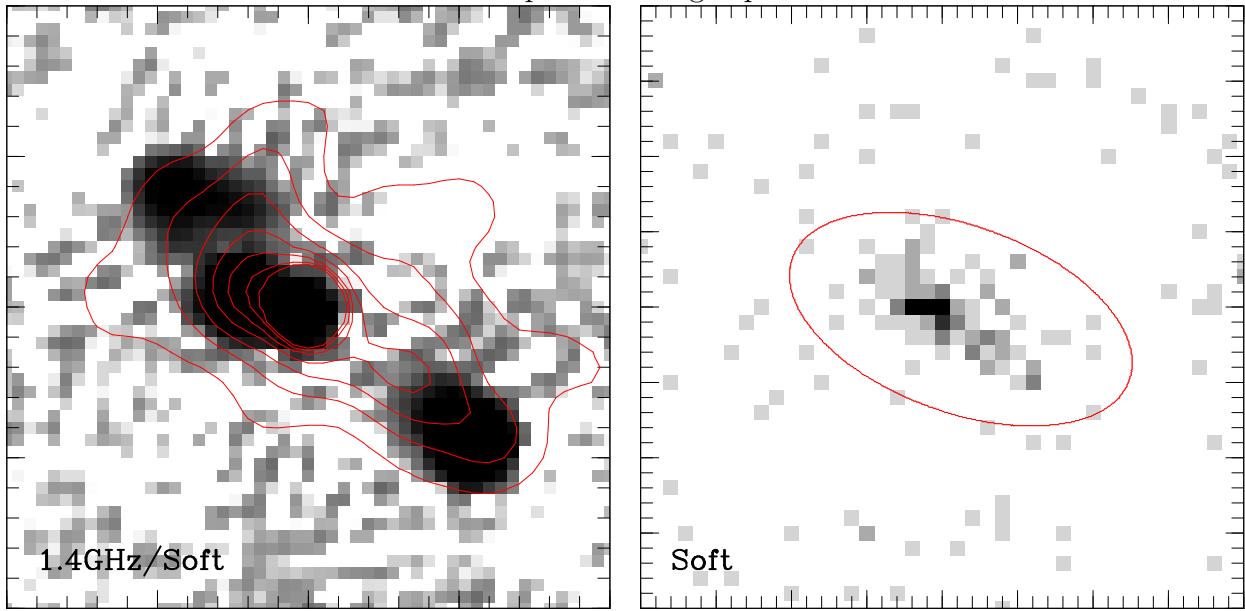}
\end{figure}

\begin{figure}
\caption{The binned spectrum of diffuse emission from 4C\,41.17
across the energy range 0.3--10\,keV (1.4--48\,keV in the restframe),
with the best fit model plotted as a solid curve and the residuals
shown in the bottom panel.  We conclude that this spectrum is well-fit
(reduced $\chi^2=1.3$) by a single slope power law with an index of
$1.29\pm 0.18$ (90\% confidence limits) and n(H{\sc i}) fixed at the
Galactic value of $1\times 10^{21}$ cm$^{-2}$.  A comparison of the
fits for the extended X-ray emission to the North-east and South-west
of the point source shows no significant difference in the spectral
properties of these two regions.  
}
\plotone{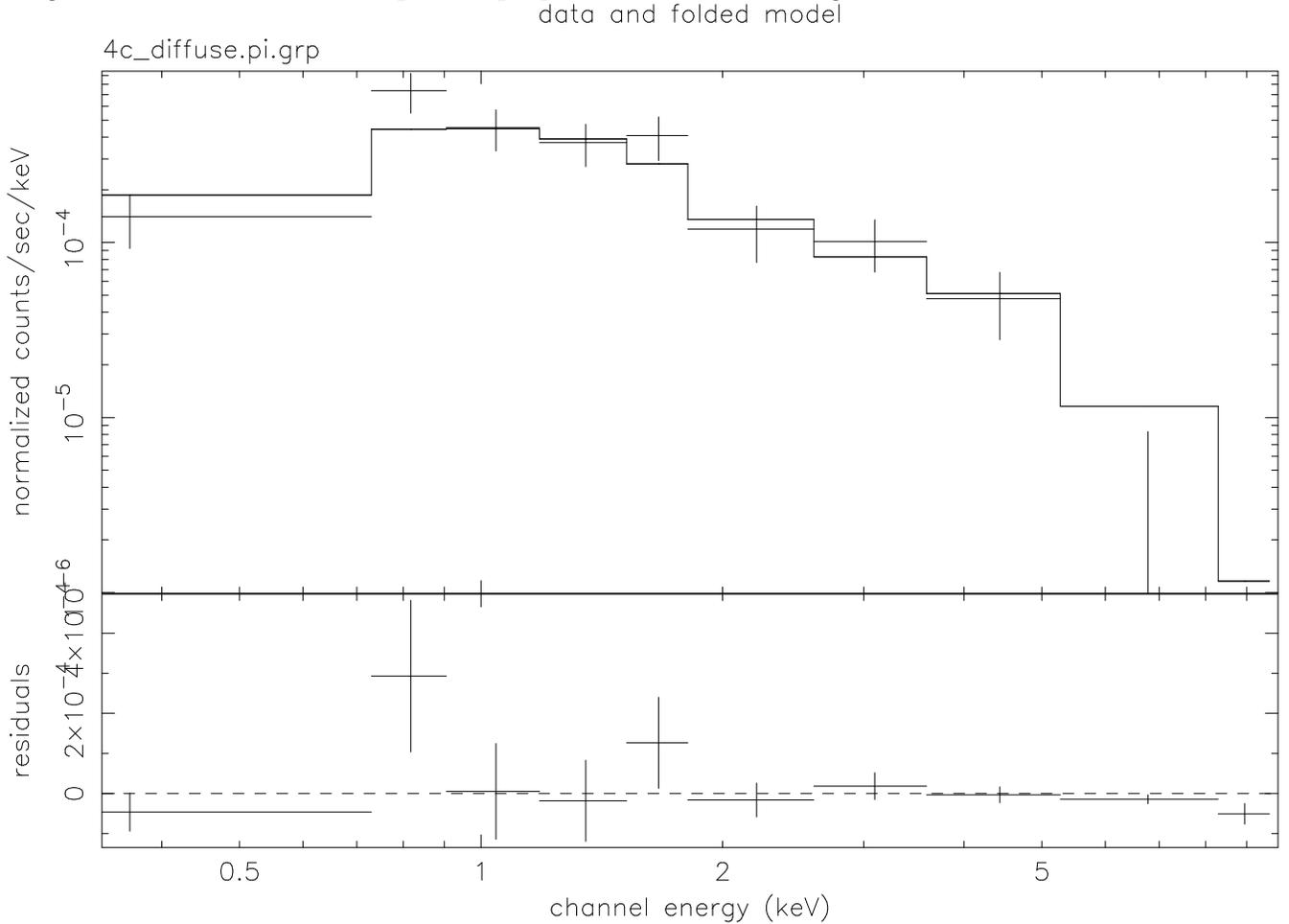}
\end{figure}

\begin{figure}
\caption{The predicted IC X-ray luminosity in the 1.44-48 keV rest frame 
(0.3-10 keV observed) is plotted versus the magnetic field strength in the 
relativistic electron population. The lowermost, light curve is the 
prediction for scattering from CMB photons only (energy density 
$\rho_{CMB}$). Curves of increasing $L_x$
correspond to the addition of a local FIR photon mean energy density in 
multiples of 
the CMB energy density: lowest curve $\rho_{FIR}=\rho_{CMB}$, middle 
$\rho_{FIR}=2\rho_{CMB}$, uppermost $\rho_{FIR}=3\rho_{CMB}$. Vertical lines
indicate the range of estimated equipartition magnetic field strength 
~\citep{carilli94}. Horizontal dashed lines indicate the allowed range of
{\em observed} X-ray luminosity in the 4C41.17 diffuse emission.
}
\plotone{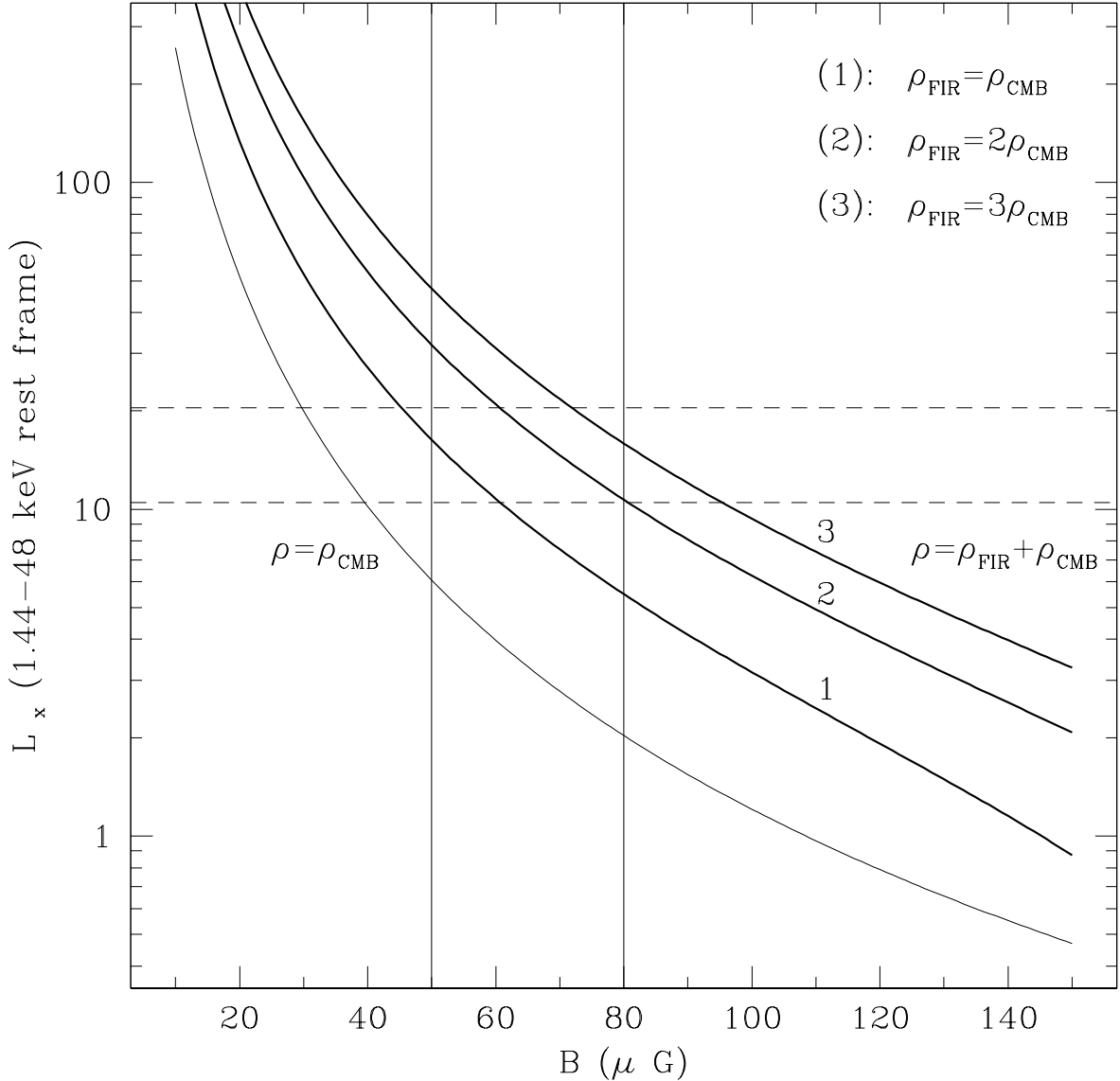}
\end{figure}

\begin{figure} \caption{ (a) The left-hand panel shows the high surface
brightness regions of the Lyman-$\alpha$ emission ~\citep{reuland03}
around 4C\,41.17 overlayed as contours (lowermost 0.22 ct per $0.21''$
pixel, linear increments of 0.3 ct per pixel) on the {\it HST} WFPC2
F702W image of the galaxy \citep{ivison00}; (b) the central panel shows
the 0.3--10\,keV adaptively smoothed {\sl Chandra} data as red contours
(lowermost 0.18 ct per $0.5''$ pixel, linear increments of 0.15 ct per
pixel) overlaid on the Lyman-$\alpha$ image of 4C\,41.17 (grayscale
0.01-2 ct per $0.21''$ pixel), the brightest regions of which are also
denoted by contours (lowermost 0.4 ct per pixel, increment 0.4 ct per
pixel); (c)  the right-hand panel shows the 1.4\,GHz radio map as
contours (red, levels 0.3, 1.9, 10.9, 21.2, 37.5, 63.3, 104.2 mJy per
pixel)  overlayed on the Lyman-$\alpha$ image (as in b)).  All panels are
$20''\times 20''$ (142\,kpc at $z=3.8$) and have North top and East to
the left.  We draw particular attention to the apparent correlation
between structures in the Lyman-$\alpha$ and X-ray wavebands shown in the
central panel.  For example, the Northern arm of the crescent-like
Lyman-$\alpha$ feature at [2$''$,$-$2$''$] (relative to the center of the
galaxy) appears correlated with a ridge in the X-ray emission, and to
some extent with the 1.4\,GHz radio emission. Similarly, in the
North/North-West of the halo three spur-like Lyman-$\alpha$ features are
seen, with matching features in the outermost X-ray emission.  These
X-ray features are also evident in the both the soft (0.5--2\,keV,
2.4--9.6\,keV rest frame) and hard (2--10\,keV, 9.6--48\,keV rest frame)
bands in Figure 1.}

\plotone{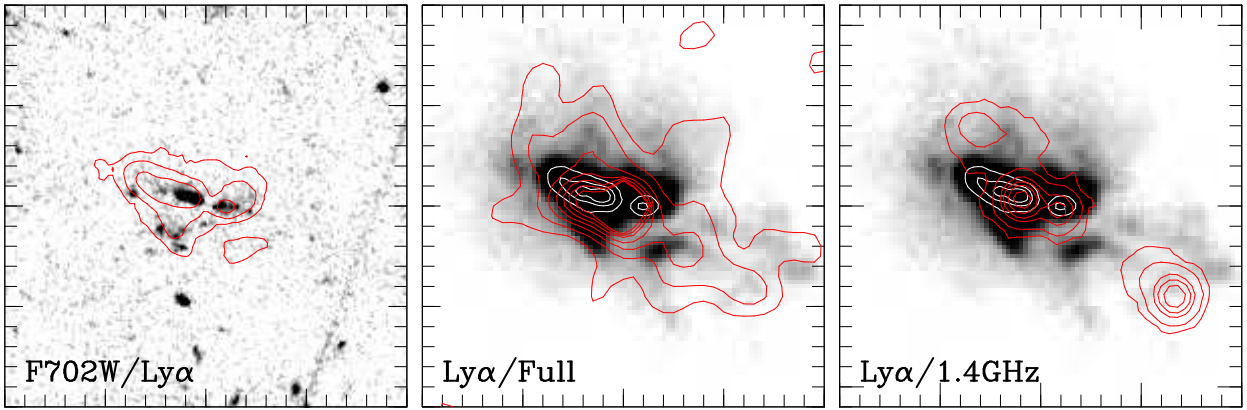}
\end{figure}

\end{document}